\documentclass[useAMS,usenatbib]{mnras}
\usepackage{graphicx,fleqn,natbib,amssymb}
\usepackage{epstopdf}
\usepackage{xcolor}
\usepackage{hyperref}
\usepackage{adjustbox}
\usepackage{multirow}
\usepackage{appendix}
\usepackage{xcolor}

\DeclareGraphicsExtensions{.eps,.cps,.ps,.eps.gz,.ps.gz,.eps.Z}
\def \ergsec{\hbox{erg s$^{-1}$}}

\def \apjl{ApJL}
\def \apjs{ApJ Suppl.}
\def \aap{A\&A}

\usepackage{dcolumn}
\usepackage{pdflscape}
\usepackage{dblfloatfix}
\usepackage{rotating}
\usepackage{longtable}
\usepackage[flushleft]{threeparttable}
\usepackage{tablefootnote}

\newcommand\Eiso{$E_{iso}$}
\newcommand{\Tn}{$T_{90}$}

\newcommand{\Rsp}{$R_{sp}$}

\newcommand{\Lg}{$L_{\mathrm{G,10s}}$}
\newcommand{\Ag}{$\alpha_{\mathrm{G,avg>10\,s}}$}
\newcommand{\Lx}{$L_{X,200s}$}
\newcommand{\Ax}{$\alpha_{X,avg>200s}$}
\newcommand{\Lo}{$L_{O,200s}$}
\newcommand{\Ao}{$\alpha_{O,avg>200s}$}

\usepackage{cleveref}

\crefformat{section}{\S#2#1#3} 
\crefformat{subsection}{\S#2#1#3}
\crefformat{subsubsection}{\S#2#1#3}

\begin{document}
\title[GRB GeV correlation]{Evidence for a luminosity-decay correlation in GRB GeV light curves}
\author[Hinds et al.]{K. R. Hinds$^{1,2}$, S. R. Oates$^{1,3}$, M. Nicholl$^{1,4}$, J. Patel$^{1}$, N. Omodei$^{5}$, B. Gompertz$^{1}$, 
\newauthor J. L. Racusin$^{6}$ and G. Ryan$^{7}$ \\
$^{1}$ School of Physics and Astronomy \& Institute for Gravitational Wave Astronomy, University of Birmingham, B15 2TT, UK\\
$^{2}$ Astrophysics Research Institute, Liverpool John Moores University, Liverpool Science Park, 146 Brownlow Hill, Liverpool L3 5RF, UK \\
$^{3}$ Department of Physics, Lancaster University, Lancaster, LA14YB, UK \\
$^{4}$ Astrophysics Research Centre, School of Mathematics and Physics, Queens University Belfast, Belfast BT7 1NN, UK \\
$^{5}$ W. W. Hansen Experimental Physics Laboratory, Department of Physics, Stanford University, Stanford, CA 94305, USA \\
$^{6}$ Astrophysics Science Division, NASA Goddard Space Flight Center, 8800 Greenbelt Rd, Greenbelt, MD 20771, USA \\
$^{7}$ Perimeter Institute for Theoretical Physics, Waterloo, Ontario N2L 2Y5, Canada \\
}

\date{Accepted...Received...}
\maketitle

\begin{abstract} 
Correlations between intrinsic properties of gamma-ray burst (GRB) light curves provide clues to the nature of the central engine, the jet, and a possible means to standardise GRBs for cosmological use. Here we report on the discovery of a correlation between the intrinsic early time luminosity, \Lg, measured at rest frame 10s, and the average decay rate measured from rest frame 10s onward, \Ag, in a sample of 13 {\it Fermi} Large Array Telescope (LAT) long GRB light curves. We note that our selection criteria, in particular the requirement for a redshift to construct luminosity light curves, naturally limits our sample to energetic GRBs. A Spearman's rank correlation gives a coefficient of -0.74, corresponding to a confidence level of 99.6\%, indicating that brighter afterglows decay faster than less luminous ones. Assuming a linear relation with $\log($\Lg$)$, we find \Ag$ = -0.31_{-0.09}^{+0.12}\log($\Lg$) + 14.43_{-5.97}^{+4.55}$. The slope of $-0.31$ is consistent at 1$\sigma$ with previously identified correlations in the optical/UV and X-ray light curves.  We speculate that differences in the rate at which energy is released by the central engine or differences in observer viewing angle may be responsible for the correlation.

\end{abstract}


\section{Introduction}
\label{intro}

Gamma-ray bursts (GRBs) are collimated relativistic jets, launched either by the core collapse in rapidly rotating massive stars (long GRBs; LGRBs), or the mergers of compact object binaries (short GRBs; SGRBs). Their observed emission comprises of two phases: initial short-lived gamma-ray emission in the range keV-MeV, known as the prompt emission, quickly followed by longer-lived emission, known as the afterglow, observed across the electromagnetic spectrum from TeV to radio \citep{Sari_1998, MAGIC_2019a, HESS_2021}. In the standard GRB fireball model, the prompt emission originates from internal shocks that take place inside the relativistic jet between shells of materials moving at different speeds, whilst the afterglow emission is created via external shocks when the jet collides with the surrounding circumstellar medium \citep[e.g.][]{Meszaros_1997, Zhang_2004, Zhang_2006}.

Sample studies of GRBs have led to the discovery of correlations linking the properties of prompt and afterglow emission, which provide invaluable insight in to the mechanisms common to all GRBs; see \cite{Dainotti_2018} for a review on various correlations. A correlation of particular interest is that found between the luminosity and average decay rate discovered in the optical/UV and X-ray afterglow light curves \citep{Oates_2012, Racusin_2016}; see also earlier work \citep{Kouveliotou_2004, Boer_2000}. The correlation, known as the luminosity-decay correlation, indicates that the more luminous light curves decay faster than their less luminous counterparts. In the case of the optical/UV afterglows, the correlation was found in a sample of 48 LGRBs and for the X-ray afterglows, it was found in 237 LGRBs\footnote{no evidence for a correlation was found in the sample of 9 X-ray SGRB light curves}. A Spearman's rank correlation was run for both studies, in the case of the optical/UV light curves, the rank coefficient, \Rsp, was determined to be $-0.58$ and the probability of the null hypothesis to be $p< 1\times10^{-5}$ \citep{Oates_2012}. For the X-ray light curves, $R_{sp} = -0.59$ and $p \ll 1\times10^{-6}$ was found \citep{Racusin_2016}. The correlation in the optical/UV and X-ray indicates that the afterglow light curves of GRBs can be described by one unifying model regardless of the detailed and varied temporal behaviour of individual LGRBs \citep{Oates_2015}. 

Observations by the {\it Fermi} Large Area Telescope (LAT) has revealed GeV light curves to have a power-law decay that extends beyond the end of the prompt emission \citep[e.g.]{Nava_2018}. These GeV light curves are likely a combination of the prompt emission and afterglow emission, with the early light curve dominated by internal shock processes (prompt emission) and the late time light curves dominated by external shock processes (afterglow) \citep[e.g][]{Nava_2018}. \cite{Panaitescu_2017} examined the $>100$ MeV flux light curves from the first \textit{Fermi}-LAT GRB catalogue \citep{Ackermann_2013b} and an additional 14 well monitored GRBs. They divided the sample into fast decaying events ($\alpha<-1.2$) and slow decaying events ($\alpha>-1.2$), finding that the light curves converged at late times and that the faster decaying events were brighter, suggesting a correlation between brightness and decay rate at high energies within the observer frame light curves.

In this paper, we expand this analysis and test if the luminosity-decay correlation found in optical/UV and X-ray, is also found at GeV energies. We construct our sample using the GeV light curves observed by the \textit{Fermi}-LAT contained in the 2nd LAT GRB catalogue \citep{Ajello_2019}. In \cref{Data_Analysis} we discuss the sample of GRBs, the fitting procedures used to measure the luminosity and decay rate, and the linear regression method performed to define the relationship. The results of this analysis are presented in \cref{Results} with the discussion and conclusions in \cref{Discussion} and \cref{Conclusions} respectively. All uncertainties throughout this paper are quoted at 1$\sigma$. Throughout, we assume the Hubble parameter H$_0 = 70\;{\rm km s}^{-1}\;{\rm Mpc}^{-1}$ and density parameters $\Omega_{\Lambda}= 0.7$ and $\Omega_m= 0.3$. 

\section{Data Analysis}
\label{Data_Analysis}
\subsection{The sample}
\label{Sample}
We obtained the \textit{Fermi}-LAT 100\,MeV-100\,GeV flux light curves from the 2nd LAT GRB catalogue \citep{Ajello_2019}. The catalogue contains 219 light curves, comprising 21 SGRBs and 198 LGRBs; SGRBs release 90\% of the prompt energy within 2s (\Tn$<2$s), and LGRBs release 90\% of the prompt energy on timescales $>2$s (\Tn$>2$s). Of these, we selected those that had measured spectroscopic redshifts taken from the 2nd LAT GRB catalogue \citep{Ajello_2019}. This criteria results in a sample of 40 GRBs, one of which we further exclude as it is the only SGRB with redshift -- in addition the X-ray and optical studies found the correlation exclusively in LGRBs -- leaving us with a sample of 39 GRBs. In section \cref{Selection Effects} we discuss how requiring spectroscopic redshifts may introduce selection effects. However, this study relates to the intrinsic luminosity in the rest frame, thus we require accurate redshifts to move from the observer frame to the rest frame.

In the following we measure the luminosity and decline rate of the light curves using a simple power-law. When fitting so that the number of data points is greater than the free parameters, we impose an additional criterion that the light curve must have at least 3 data points included in the fit. 
This criterion reduces the final sample to 14 LGRBs.

\subsection{Luminosity Light curves}
\label{Light Curves}
We define the start time of our light curves, $T_0$, as the end time of the GBM \Tn\ parameter, consistent with the procedure of \citet{Oates_2012} and \citet{Racusin_2016}\footnote{Note \citet{Oates_2012} and \citet{Racusin_2016} used {\it Swift} Burst Alert Telescope (BAT) detected GRBs and therefore use the end time of the \Tn\ parameter measured by {\it Swift}/BAT}. We then converted each of the GeV flux light curves into the rest frame. All times were divided by a factor 1+$z$ and the luminosity defined by 

\begin{equation}
    L(t)= F_{\nu}(t) \times 4\pi D_{l}^2 (1+z)^{\beta - 1},
    \label{eq:k_correction}
\end{equation}
where $D_l$ is the luminosity distance, $z$ is the redshift and $\beta$ is the spectral index of the GRB. The temporal and spectral indices, $\alpha$ and $\beta$, are given by the expression $F(t,\nu)\propto t^{\alpha}\nu^{\beta}$. A photon index, $\Gamma$, was provided for each flux point in the LAT GRB catalogue, where $\Gamma=  \beta +1$. For this analysis we take $\Gamma$ to be the average of the values computed for each GRB -- these are listed in Table \ref{table:grb_all}. 

\subsection{Intrinsic Early Time Luminosity \& Power-law Fits}
\label{Early Luminosity}
We first define a time at which we measure the luminosity and from this, we fit a power-law to the rest of the light curve to measure the average decay index. The luminosity-decay correlations from \citet{Oates_2012} and \citet{Racusin_2016} were exclusively found in the afterglow regime and not the prompt. For the GeV sample, we therefore need to select a time that passes through as many GRB light curves as possible, is early (to maximise the dynamic range in luminosity), but not too early so as to avoid the very earliest behaviour which exhibits prompt emission features \citep[e.g.][]{Ackermann_2011,Nava_2018}; we chose this time to be $10$s.

To measure the 100\,MeV-100\,GeV luminosity at $10$s, \Lg, we fit a power law to the data within the time range $\log(T/{\rm s}) = 1 \pm 30\%$; corresponding to fitting data points within $\sim5 - 20$s. A second power-law is fit to the data from $10$s onward, to measure the decay rate \Ag. These fits are performed using the python module \texttt{lmfit}. 

\begin{figure}
    \centering
    \includegraphics[width=1\linewidth]{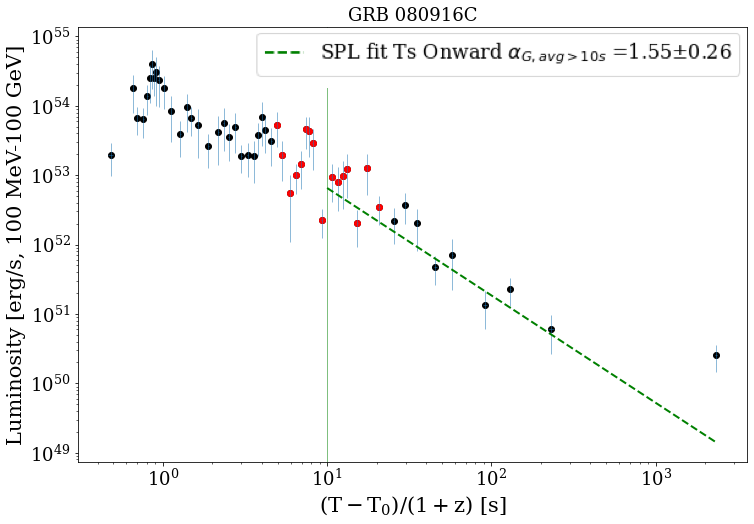}
    \caption{Light curve of GRB 080916C. The green dashed line shows the simple power-law (SPL) fit to the data from 10s onward. The red data points in the range $\log(T/{\rm s}) = 1 \pm 30\%$ were used in a separate fit to calculate \Lg.}
    \label{fig:916C LC}
\end{figure}

\begin{figure}
    \centering
    \includegraphics[width=1\linewidth]{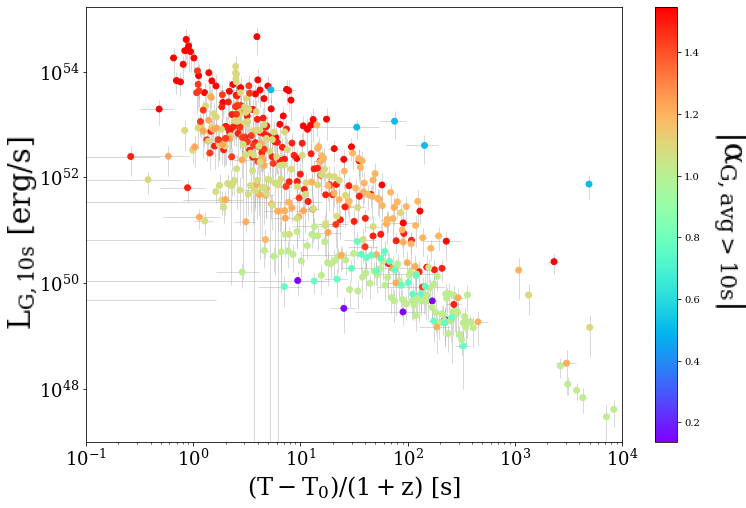}
    \caption{Final 14 LGRB light curves, colourmapped according to their absolute values of \Ag. The GeV GRB afterglow light curves appear to cluster more tightly in luminosity at later times. The colour mapping suggests that the more luminous the GRB the faster its decay. The exception is GRB 170405A, which is significantly brighter at late times compared to the rest of the sample.}
    \label{fig:colourmap}
\end{figure}

By fitting a simple power-law (SPL) to the light curves from restframe 10s onwards, we are probing the average rate that light curves decay, rather than the detailed underlying behaviour. It is well established that the X-ray afterglow light curves display a canonical behaviour, a power law decay with one or more light curve segments \citep[e.g.][]{Zhang_2006}. In \cite{Racusin_2016}, a correlation was found between the X-ray luminosity at restframe 200s, $L_{{\rm X,200s}}$, and the average decay index. They also tried correlating individual light curve segments with $L_{{\rm X,200s}}$ to test whether one segment was more significant than the others. However, they found that the correlation was not significant for any of the individual segments of the canonical light curve with $L_{{\rm X,200s}}$, indicating the importance of the average decay measure. 

We also investigated the affect of measuring the correlation at times later than rest frame 10s, e.g measuring the luminosity at 20s, 30s and 40s, and also the average decay index using data from the same time the luminosity is measured and beyond. In each instance, the range in measured luminosities at these times decreases making it increasingly more difficult to recover a correlation. In addition, the average number of data points per light curve decreases as we consider time ranges that start later. At 10s onwards, the average number of data points per GRB light curve for this sample is $\sim18$ while from 40s onwards the average number of data points is $\sim12$ but the coverage is not consistent across the sample. Conversely, using times earlier than 10s increases the risk of sampling a larger contribution from the prompt phase or the subsequent transition from prompt to afterglow.

\subsection{Determining a Relationship}
\label{GeV Correlation}
To determine if luminosity is correlated with the decay rate we perform a Spearman's rank test, which is a non-parametric measure of the strength and direction of any correlation. We also performed a `partial' Spearman's rank test, which takes into account the effect of a third parameter; this was to determine if systematic effects due to redshift could be responsible for the correlation. The results of both the standard and `partial' Spearman's rank analysis are presented in Table \ref{table:final_correlation}. Following on, linear regression was performed using the \texttt{ODR} python module which defined the relationship between the two parameters; the python \texttt{ODR} linear regression results were compared with the IDL routine \texttt{FITEXY}, which was used by \cite{Oates_2012} and \cite{Racusin_2016}, and regression parameters from both methods were found to be consistent within 1$\sigma$. The errors on the Spearman's rank and linear regression were calculated using Monte-Carlo methods. \cite{Curran_2014} discussed whether a Bootstrapping method or Resampling each point within its uncertainties is optimal for calculating errors; for this analysis, we use both methods individually and also use a combination of the two. In each case, we ran the Monte Carlo simulations for $10^5$ trials. In an attempt to be thorough with our error analysis, we favour the combination method which includes Bootstrapping and then Resampling - these are the errors presented in Table \ref{table:final_correlation}.  

\section{Results}
\label{Results}
Examining the distribution of light curves in Fig.~\ref{fig:colourmap}, we see the light curves cluster. Note the greatest spread in luminosity is at early times and the distribution appears to become narrower with time. In addition to this, when colouring the luminosity light curves by their average decay rate we see a colour gradient, which serves as visual confirmation of the correlation (that the more luminous light curves decay faster). There is one outlier, GRB 170405A, that appears to be offset at a higher luminosity compared to the other GRBs. 

We first perform a Spearman's rank test on the entire sample of 14 light curves. This results in a correlation coefficient of $-0.44\pm0.31$ and a p-value of $1.14\times10^{-1}$. With the large error on the Spearman's rank coefficient, we cannot claim a correlation between the two parameters. However, we note the exceptionally flat light curve of GRB 170405A that stands out in Fig.~\ref{fig:colourmap}, and suggests this may have a different emission origin compared to the rest of the sample or that the spectral index used in Eq. \ref{eq:k_correction} is inaccurate (see section \ref{Discussion}). We, therefore, performed the Spearman's rank test after removing this GRB from the sample. In this case we find a significant negative correlation, with a coefficient of $-0.74\pm0.19$ and p-value of $4.11\times10^{-3}$. For the `partial' Spearman's rank, we found a coefficient of $-0.44$ and p-value of $1.37\times10^{-1}$.

\begin{figure}
\centering
\includegraphics[width=1\linewidth]{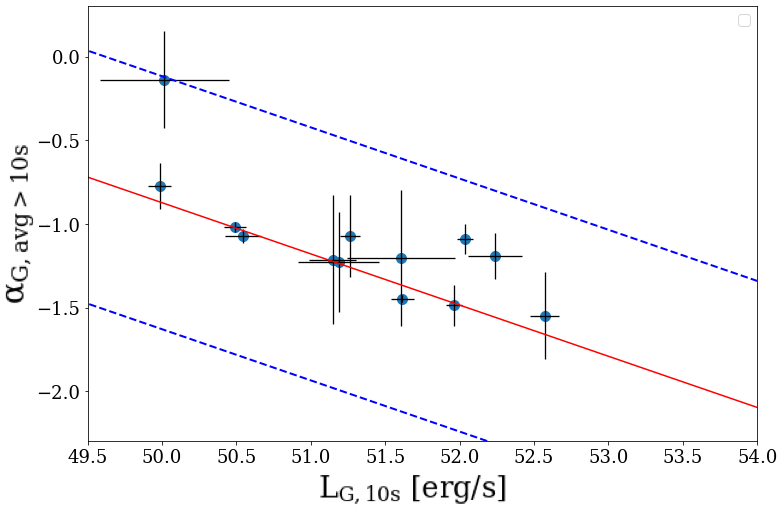}
\caption{The GeV average decay rate from restframe 10s onwards against luminosity measured at restframe 10s. The solid red line is the best fitting linear regression relationship and the blue dashed lines represent the $3\times$\,RMS (root-mean-square) variation. The Spearman's rank coefficient is $-0.74\pm0.19$ and the probability of the null hypothesis (no correlation) is $4\times10^{-3}$. We measure a linear relationship  \Ag$ = (-0.31_{-0.09}^{+0.12}$) $\log($\Lg$) + 14.43_{-5.97}^{+4.55}$. }
\label{fig:lin reg}
\end{figure}

Fig.~\ref{fig:lin reg} shows the luminosity vs decay rate for the final 13 GeV light curves. We also performed a linear regression which gives a relationship \Ag$ = (-0.31_{-0.09}^{+0.12}$) $\log($\Lg$) + 14.43_{-5.97}^{+4.55}$. This line is over-plotted in Fig. \ref{fig:lin reg}. Table \ref{table:final_correlation} gives the results of the Spearman's rank and linear regression analyses. 

\begin{table}
  \centering
    \begin{tabular}{|c|c|c|c|c|}
        \hline
        GRB & $z$  &$\Gamma$ & \Ag & \Lg (\ergsec{}) \\ \hline   

        080916C & 4.35 & $-2.60\pm0.54$ & $-1.55^{+0.26}_{-0.26}$ & $3.73_{-0.99}^{+0.92}\times10^{52}$ \\ 
        
        090323 & 3.57  & $-2.29\pm1.11$ & $-1.21_{-0.39}^{+0.42}$ & $4.05_{-3.13}^{+2.80}\times10^{51}$ \\ 
        
        090902B & 1.82  & $-1.96\pm0.18$ & $-1.49_{-0.13}^{+0.12} $& $9.16_{-1.06}^{+1.05}\times10^{51}$ \\ 

        090926A & 2.11 & $-2.21\pm0.46$ & $-1.09_{-0.11}^{+0.07}$ & $1.09_{-0.15}^{+0.15}\times10^{52}$ \\ 
        
        91003   & 0.90  & $-1.88\pm0.26$ & $-0.14_{-0.27}^{+0.31}$ & $1.03_{-1.03}^{+1.03}\times10^{50}$ \\ 

        110731A & 2.83 & $-2.33\pm0.64$ & $-1.23_{-0.31}^{+0.28}$ & $1.54_{-0.71}^{+0.78}\times10^{51}$ \\ 
        
        130427A & 0.34  & $-2.02\pm0.24$ & $-1.02_{-0.03}^{+0.03}$ & $3.10_{-0.35}^{+0.35}\times10^{50}$ \\ 
        
        131108A & 2.40  & $-2.64\pm0.69$ & $-1.45_{-0.02}^{+0.02}$& $4.12_{-0.80}^{+0.75}\times10^{51}$ \\ 
        
        141028A & 2.33  & $-2.43\pm0.48$ & $-1.07^{+0.23}_{-0.26}$ &$1.84_{-0.31}^{+0.31}\times10^{51}$ \\ 
        
        160509A & 1.17  & $-2.45\pm1.78$ & $-1.21_{-0.36}^{+0.42}$ &$1.40_{-0.33}^{+0.32}\times10^{51}$ \\ 
        
        170214A & 2.53  & $-2.47\pm0.58$ & $-1.19_{-0.14}^{+0.14}$ & $1.72_{-0.88}^{+0.81}\times10^{52}$ \\ 
        
        170405A & 3.51  & $-5.58\pm2.85$ & $-0.51_{-0.11}^{+0.09}$ & $2.58_{-2.58}^{+2.58}\times10^{53}$ \\ 
        
        180720B & 0.65  & $-2.26\pm0.33$ & $-0.77_{-0.14}^{+0.14}$ & $9.62_{-0.63}^{+0.63}\times10^{49}$ \\ 
        
        190114C & 0.42  & $-2.10\pm0.60$ & $-1.07_{-0.04}^{+0.04}$ & $3.49_{-2.21}^{+2.78}\times10^{50}$ \\ \hline

    \end{tabular}
  \caption{ The sample parameters: GRB, redshift (provided in the 2nd LAT GRB catalogue; \citealt{Ajello_2019}), mean photon index ($\Gamma$), and \Ag~and \Lg~which are the average decay rate from restframe 10s onward and the intrinsic early time luminosity calculated at restframe 10s. Errors are given at $1\sigma$ confidence.}
  \label{table:grb_all}
\end{table}

\begin{table*}
    \begin{tabular}{|c|c|c|c|c|c|c|c|c|}
    \hline
        \multicolumn{2}{|c|}{Parameters}  &  
        \multirow{2}{2cm}{Spearman's Rank} &
        \multirow{2}{1.5cm}{Null Hypothesis}&
        \multirow{2}{2.5cm}{Partial Spearman's Rank}&
        \multirow{2}{1.5cm}{Null Hypothesis}&
        \multicolumn{2}{c|}{Linear Regression}&
        \multirow{2}{1cm}{No. GRBs}\\ 
        x-axis & y-axis &  & & & &slope& intercept &\\ \hline  
            \Lg$^{(1)}$ & \Ag & $-0.44\pm0.33$ & $1.14\times10^{-1}$ & -0.13 & $6.81\times10^{-1}$ & $-0.34_{-0.21}^{+0.18}$& $16.42_{-9.10}^{+10.70}$& 14 \\
            
            \Lg$^{(2)}$ & \Ag & $-0.74\pm0.19$ & 4.11$\times10^{-3}$ & -0.45 & $1.37\times10^{-1}$ & $-0.31_{-0.09}^{+0.12}$& $14.43_{-5.97}^{+4.55}$&13 \\ 
            
            \Lg$^{(3)}$ & \Ag & $-0.74\pm0.19$ & 4.11$\times10^{-3}$ & -0.46 & $1.37\times10^{-1}$ & $-0.31_{-0.05}^{+0.05}$& $1.06_{-0.04}^{+0.04}$&13 \\

            \Lo$^{(4)}$ & \Ao & $-0.58\pm$ 0.11 & $1.90\times10^{-5}$ & -0.50 & $2.85\times10^{-4}$ & $-0.28_{-0.04}^{+0.04}$ & $7.72_{-1.31}^{+1.31}$ &48 \\
      
            \Lx$^{(5)}$ & \Ax & $-0.59 \pm$ 0.09 & $8.03\times10^{-8}$ & -0.63 & $1.58\times10^{-6}$ & $-0.27_{-0.04}^{+0.04}$ & $6.99_{-1.11}^{+1.23}$ &237 \\ 
            \hline
    \end{tabular}
    \caption{This table contents include the x and y axis parameters used in the Spearman's rank tests, the Spearman's rank coefficient and probability of null hypothesis, `partial' Spearman's rank and the corresponding probability of null hypothesis, linear regression slope and intersect and the number of GRBs used in each run. $^{(1)}$ denotes the run that included 170405A, $^{(2)}$ denotes the run excluding 170405A,$^{(3)}$ denotes the results with luminosity values normalised by $10^{51}$, $^{(4)}$ found in \protect\cite{Oates_2012} and $^{(5)}$ found in \protect\cite{Racusin_2016}.}
    \label{table:final_correlation}
\end{table*}

\section{Discussion}
\label{Discussion}
Overall, we have shown that a correlation exists between the intrinsic brightness of GeV light curves and their average decay rate. In the following section, we discuss the origin of the GeV emission and whether it is appropriate to exclude 170405A. We then compare our results with the correlation found in the optical/UV and X-ray samples.

\subsection{GeV Emission Mechanisms}
\label{GeV Emission Mechanism}
The GeV emission observed by \textit{Fermi}-LAT is thought to be a combination of emission processes from the internal and external shocks that dominate at different times during the evolution of the GeV emission \citep[see][for a review]{Nava_2018}. At early times, the GeV light curves often correlate with the flux observed at MeV energies \citep{Tang_2017}, while spectrally they can either be fit with an extension of the power-law from the MeV energy range or have an additional power-law component \citep{Maxham_2011,Panaitescu_2017, Ajello_2019, Fraija_2020}, for which the origin may be synchrotron self-Compton (SSC) emission \citep{Ackermann_2011, Nava_2018}. 

The external shock emission thought to produce the afterglow is unable to reproduce the GeV flux at very early times \citep{Maxham_2011}. Instead, the early GeV emission is expected to be dominated by synchrotron and SSC emission components, originating from the internal shock that drives the prompt emission \citep{Maxham_2011, Peer_2012, Fraija_2020}. Following the prompt emission is a regime labelled the ‘GeV extended emission’ \citep[e.g.][]{Ackermann_2014}. From this point onward, the temporal behaviour in the GeV band is a power-law decay similar to the canonical X-ray afterglow light curve \citep{Nousek_2006, Zhang_2006}. This emission is thought to be dominated by synchrotron radiation \citep[e.g.][]{Ackermann_2011, Kumar_2010, Toma_2011, Feng_2011,Ajello_2018, Nava_2018, Maxham_2011, Beniamini_2015, Ajello_2019, Tak_2019}. Though for some GRBs, particularly those with photons $>10$\, GeV, SSC emission can explain the observed emission \citep{Fraija_2022}. In the case of GRB 221009A, a narrow jet, $\sim0.8^{\circ} $, and SSC of electrons in the external shock has been suggested to explain observations of TeV photons from the afterglow \citep{LHAASO_2023}.

By excluding GRB 170405A from our sample, the outlier in our luminosity distribution Fig.~\ref{fig:colourmap}, we find a strong correlation between the brightness of the GeV luminosity light curves and their average rate of decay. This prompted us to investigate why 170405A is an outlier. We searched the literature to determine if GRB 170405A is produced by different emission processes compared to the other GRBs. \cite{Tak_2019} compared the temporal and spectral behaviour of the GeV extended emission to the synchrotron external shock model and showed that most GRBs could be explained by this model. This analysis included GRBs 080916C, 090323A, 090926A, 091003A, 110731A, 130427A, 131108A, 141028A, 160509A, 170214A, 170405A \& 180720B from our sample. However, using multi-wavelength observations, \cite{Arimoto_2020} found that the GeV emission from 170405A could not be produced by the same component as the optical/UV emission and that the GeV emission must be produced by either a different external shock component or more likely produced by internal processes. This suggests that it is important to examine multi-wavelength observations in order to confirm the origin of the GeV emission. 

Further exploring the literature, we find that the external forward shock model is shown to reproduce the late GeV emission of the light curves of all the other GRBs in our sample \citep{Kumar_2010, Swenson_2010, Ackermann_2011, Barniol_2011, Feng_2011,Maxham_2011, Piron_2011, Toma_2011,  Ackermann_2013,Fan_2013,  Kouveliotou_2013, Liu_2013, Wang_2013, Ackermann_2014, Maselli_2014, Perley_2014, Vestrand_2014, Beniamini_2015, Fraija_2015, Burgess_2016, Lu_2017, Panaitescu_2017, Tam_2017, Nava_2018, Ajello_2019, Fraija_2019, Ronchi_2020, Fraija_2021, Joshi_2021}. However, the picture is not completely clear cut, as some authors invoke additional components to produce some or all of the late time LAT emission \citep{Liu_2013,Tam_2017, Duan_2019, Wang_2018}. For instance, SSC emission may better explain the observed LAT emission \citep{Fraija_2022}, particularly for those GRBs with photons $>10$\,GeV. Inverse Compton (IC) could also explain the highest energy GeV photons in GRBs such as GRB 130427A, 160509A, 180720B \citep{Fan_2013, Liu_2013, Tam_2013, Wang_2013, Ackermann_2014, Tam_2017, Fraija_2019}. While late GeV light curves from LAT are typically dominated by low-energy photons \citep[e.g.][]{Ackermann_2011,Nava_2018}, likely produced by the external forward shock model, other emission components such as SSC may contribute, particularly producing the highest energy photons.

Examining Table \ref{table:grb_all}, we note that the photon index for 170405A, $-5.58\pm2.85$, is especially large when compared to the mean of the sample, $-2.8\pm0.57$, which may account for why this GRB is an outlier. In the LAT catalog paper \citep{Ajello_2019}, the photon index for 170405A, determined between 18 and 868s, is $-2.8\pm 0.3$. For our analysis we have used the average photon index of the entire LAT light curve of GRB 170405A, provided in the LAT catalog and we note that the earliest spectral bins, with times $<18$s, have values of the photon index $\ll-2.8$. \cite{Arimoto_2020} also report photon indices for the LAT data. In two time intervals 310-560s and 589-1000s (observer frame), they report LAT photon indices of $-1.88\pm0.33$ and $-2.36\pm0.50$, which are consistent within 1.40 and 0.58 $\sigma$ respectively, of the average photon index of our sample. Therefore to test if this photon index of $-5.58\pm2.85$ is anomalous, we assumed the mean of our sample as the photon index of 170405A, recomputed its luminosity light curve and then reran the analysis. We found that the light curve of 170405A decreased in luminosity by approximately two orders of magnitude. It no longer appears as an outlier and is consistent in luminosity with the other GRBs in the sample. Rerunning the correlation gives a result consistent with that found in the sample of 13 GRBs - the slope of the linear regression being consistent within 1$\sigma$ of their respective errors. Since it is unclear whether this GRB is an outlier due to physical differences in the origin of this particular GRB or uncertainty in the photon index measurement, we will continue to discuss the GeV luminosity-decay correlation excluding GRB 170405A.

\subsection{Comparison with Previous Correlations}
\label{Comparison}
Due to very few of the GeV, optical and X-ray light curves overlapping at restframe 10s or 200s, we are unable to directly compare the luminosity-decay correlation found at GeV energies using the same time as that for the optical and X-ray. However, we can compare the parameters and strength of the correlation derived using data covering the different time ranges. In Table \ref{table:final_correlation}, we provide the results of the optical/UV and X-ray correlation analyses presented in \cite{Oates_2012} and \cite{Racusin_2016} - we also provide a more physical interpretation of the GeV correlation with the luminosities normalised by $\times10^{51}$. Comparing the results of our GeV sample with the optical/UV results, we find the linear regression slope and intercept are consistent within 0.27$\sigma$ and 1.27$\sigma$, respectively. For the X-ray study, we find the linear regression slope and intercept are consistent within 0.36$\sigma$ and 1.42$\sigma$, respectively. The consistency of the correlation slopes across $10^{10}$ orders of magnitude in energy (from optical photons to GeV photons) indicates the processes producing the emission are likely to be the same mechanism and provides additional support for the GeV light to originate from an external shock, at least after rest frame 10s. 

The GeV light curves are shorter in duration and cover an earlier time range compared to the optical/X-ray with the GeV lasting $\sim10^1-10^3{\rm s}$ and the optical/X-ray lasting $\sim10^2-10^7{\rm s}$. This implies that GeV light curves have the potential to be in the fast cooling phase whilst the optical/X-ray is typically in the slow cooling regime \citep{Zhang_2006,ghi10,Ajello_2019}. \citet{Tak_2019} looked at the closure relations for the GeV extended emission of 13 out of 14 GRBs in our sample. They determined that 7 of the GRBs in our sample are consistent with the fast cooling regime with $\nu>\nu_m, \nu_c$, where $\nu_c$ is the synchrotron cooling frequency and $\nu_m$ is the synchrotron peak frequency. Four are consistent with being in the slow cooling regime with $\nu_m<\nu<\nu_c$ and two are unclassified. We split the sample based on their cooling regime and tested the correlation strength to determine whether the correlation is driven by a certain cooling regime. The Spearman rank test for fast cooling only and slow cooling only gives coefficients of -0.68 and -0.60, and p-values of 0.09 and 0.40, respectively. Although the p-values are larger due to the smaller number of GRBs involved in each correlation, the coefficients are a similar value to that found for the full value. This suggests that the luminosity-decay correlation in the GeV energy range is not affected or produced by differences in cooling regime. This is also supported by Fig 1 of \cite{Tak_2019}, which shows similar observed temporal indices for LAT light curves consistent with either fast or slow cooling regimes. 

\cite{Oates_2015}, simulate the relationships, expected between $\log L_{200{\rm s}}$ and $\alpha_{>200{\rm s}}$ and isotropic gamma-ray energy $\log$ \Eiso~from a basic afterglow model, for the optical and X-ray afterglows. They conclude that the simulations do not agree with correlations observed between $\log L_{200{\rm s}}$ and $\alpha_{>200{\rm s}}$, or $\log$ \Eiso~and $\alpha_{>200{\rm s}}$. This suggests that while a common underlying physical mechanism is consistent with producing GRBs and their optical and X-ray afterglows, regardless of their detailed afterglow light curve behaviour, a basic afterglow model has difficulty explaining all the observed correlations. Instead, the luminosity-decay correlation could be a result of different rates of energy deposition from the central engine to the surroundings; faster decays occur when the energy is deposited rapidly from the central engine, and hence produce initially more luminous afterglows \citep{Oates_2012, Oates_2015, Panaitescu_2017}. An alternative explanation may be that the jet is viewed off-axis and may be structured \citep{Oates_2012,Oates_2015}. When a jet is viewed at large angles away from the jet axis, a GRB can appear to be dimmer and decay on a longer timescale compared to GRBs that are observed close to the jet axis \citep[see also][]{Granot_2002,Rossi_2004,Ramirez_2005, Panaitescu_2008,Ryan_2020}. Structured jets have been used to explain the brightest GRB afterglows such as that of GRB~221009A \citet{OConnor_2023}.

\subsection{Possible Selection Effects}
\label{Selection Effects}
\begin{figure}
    \centering
    \includegraphics[width=1\linewidth]{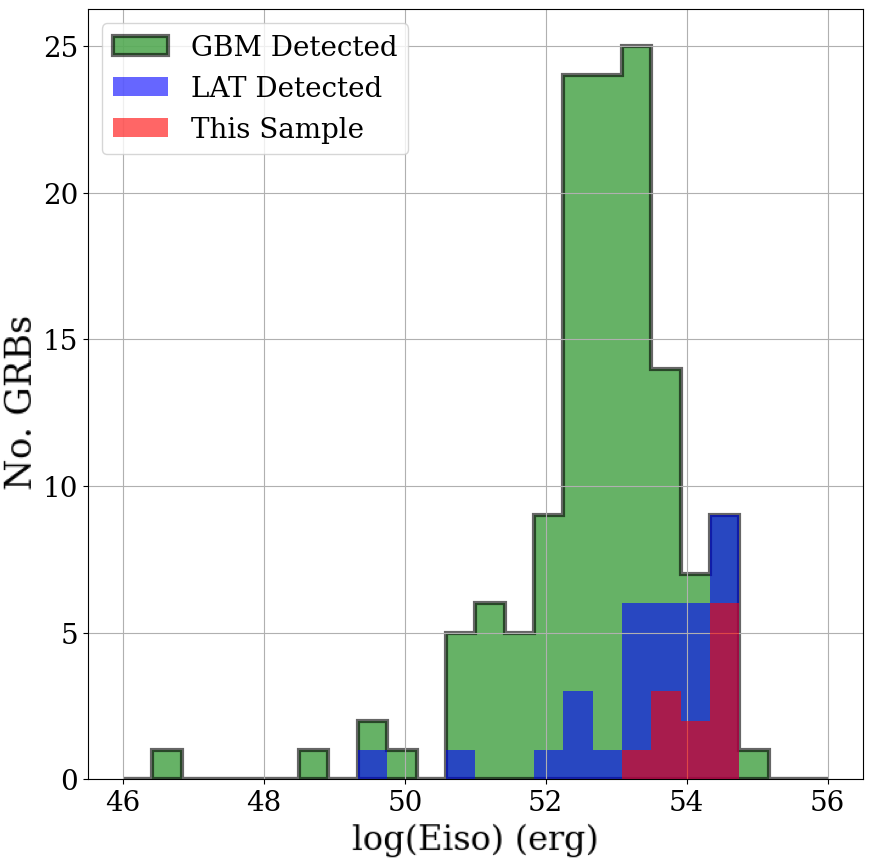}
    \caption{Isotropic gamma-ray energy, \Eiso, distribution of the GeV afterglows in this sample (red), the 2nd LAT catalogue (blue) and the GBM measured \Eiso (green; \citet{Poolakkil_2021}).}
    \label{fig:eiso_dists}
\end{figure}
Requiring a spectroscopic redshift notably reduces the number of LAT light curves in our sample. However, it is necessary to construct rest-frame light curves in order to directly compare the intrinsic brightness of different GRBs. We also work in the rest frame to be able to compare the results of this paper with previously found correlations at other wavelengths by \citet{Oates_2012} and \citet{Racusin_2016}. This redshift requirement introduces some selection biases. 
The gamma-ray emission of GRBs is in general not well localised. Unlike {\it Swift}, {\it Fermi} does not have narrow field instruments onboard and so follow-up of the GRB afterglow at longer wavelengths, which provides better positional accuracy and enables spectroscopic follow-up, occurs later than for {\it Swift} detected GRBs. This implies that spectroscopic follow-up of {\it Fermi}-LAT detected GRBs is only achievable for those that have afterglows bright enough at late times to obtain a spectrum. We attempt to quantify this selection bias, by comparing the distributions of isotropic gamma-ray energy, \Eiso, of this sample with the whole LAT catalogue, using the GBM measured \Eiso; the isotropic energy is correlated with afterglow brightness \citep[e.g.][]{Davanzo_2012, Margutti_2013, Oates_2015}. These distributions are shown in Figure \ref{fig:eiso_dists}, together with the \Eiso values of the entire GBM sample. We ran a two-sample Anderson-Darling test on our sample and the full {\it Fermi} sample to address whether they are statistically different. A bias towards brighter GRBs is apparent visually in Figure 4. A two-sample Anderson-Darling test, comparing our sample (red) to the LAT sample (blue) and to the GBM sample (green), gives p=0.07 and p=0.01 respectively. The comparison to the LAT sample is only marginally significant and likely due to the small size of our sample. However, comparison with the GBM sample is more significant and indicates we are biased towards energetic events.

\section{Conclusions}
\label{Conclusions}
We examined a sample of 13 LAT light curves to determine the relationship between the intrinsic early time luminosity, \Lg, and average decay index, \Ag. From the Spearman's rank test we found a coefficient of $-0.74\pm0.19$ and p-value $4.11\times10^{-3}$, indicating a correlation is present such that the brightest GeV light curves decay on average faster than fainter GeV light curves. A linear regression between the two parameters gives \Ag$ = -0.31_{-0.09}^{+0.12}\log($\Lg$) + 14.43_{-5.97}^{+4.55}$, consistent with optical/UV and X-ray measurements of a similar correlation to within 0.4$\sigma$ and 1.4$\sigma$ in the slope and intercept, respectively. This consistency suggests the mechanism producing the GeV luminosity-decay correlation is the same as that producing the correlation observed in the optical/UV and X-ray light curves. It suggests that they are all produced by the same emission component, further supporting the forward shock being the dominant emission mechanism of GRB GeV light curves from around rest frame 10s onward, at least for the GRBs in this sample. Due to the sample size and requirement of redshifts, we have discussed possible selection biases and how representative our sample is compared to LAT detected and GBM detected GRBs; statistical tests suggest our sample is biased towards energetic GRBs.

\section{Acknowledgments}
The \textit{Fermi} LAT Collaboration acknowledges generous ongoing support
from a number of agencies and institutes that have supported both the
development and the operation of the LAT as well as scientific data analysis.
These include the National Aeronautics and Space Administration and the
Department of Energy in the United States, the Commissariat \`a l'Energie Atomique
and the Centre National de la Recherche Scientifique / Institut National de Physique
Nucl\'eaire et de Physique des Particules in France, the Agenzia Spaziale Italiana
and the Istituto Nazionale di Fisica Nucleare in Italy, the Ministry of Education,
Culture, Sports, Science and Technology (MEXT), High Energy Accelerator Research
Organization (KEK) and Japan Aerospace Exploration Agency (JAXA) in Japan, and
the K.~A.~Wallenberg Foundation, the Swedish Research Council and the
Swedish National Space Board in Sweden.
 
Additional support for science analysis during the operations phase is gratefully 
acknowledged from the Istituto Nazionale di Astrofisica in Italy and the Centre 
National d'\'Etudes Spatiales in France. This work performed in part under DOE 
Contract DE-AC02-76SF00515.

\section{Data Availability}
The data underlying this article were obtained from the 
2nd \textit{Fermi}/LAT GRB catalogue \citep{Ajello_2019}. The data are available 
at https://heasarc.gsfc.nasa.gov/W3Browse/fermi/fermilgrb.html and https://www-glast.stanford.edu/pub\_data/1874/

\bibliographystyle{mn2e}   
\bibliography{references_v2.bib} 
\newpage

\IfFileExists{\jobname.bbl}{}
 {\typeout{}
  \typeout{******************************************}
  \typeout{** Please run "bibtex \jobname" to optain}
  \typeout{** the bibliography and then re-run LaTeX}
  \typeout{** twice to fix the references!}
  \typeout{******************************************}
  \typeout{}
 }
 
\end{document}